\documentclass[]{aa}
\usepackage[dvips]{epsfig}
\usepackage{natbib}
\bibpunct{(}{)}{;}{a}{}{,}
\begin{document}

\newcommand{\beqa}{\begin{eqnarray}}
\newcommand{\eeqa}{\end{eqnarray}}
\newcommand{\beq}{\begin{equation}}
\newcommand{\eeq}{\end{equation}}
\newcommand{\beqn}{\[}
\newcommand{\eeqn}{\]}
\newcommand{\dff}{\mathrm{d}}
\newcommand{\NN}{\mathrm{\mathsf{I}\! N}}
\newcommand{\RR}{\mathrm{\mathsf{I}\! R}}
\newcommand{\cms}{\,\mathrm{cm\,s^{-1}}}
\newcommand{\ms}{\,\mathrm{m\,s^{-1}}}
\newcommand{\kms}{\,\mathrm{km\,s^{-1}}}
\newcommand{\pderI}[2]{\frac{{\partial} #1 }{{\partial} #2 }}
\def \Dsh{1+\frac{v}{c}}

\title{A multi-method approach to radial-velocity measurement 
for single-object spectra}
\titlerunning{A multi-method approach to radial-velocity measurement}
\author{M. David \inst{1} \and
 R. Blomme \inst{2} \and
Y. Fr\'emat \inst{2} \and
Y. Damerdji \inst{3,4} \and
C. Delle Luche \inst{5} \and
E. Gosset \inst{4} \and
D. Katz \inst{5} \and
Y. Viala \inst{5}}
           
\institute{ 
Departement Wiskunde en Informatica, Universiteit Antwerpen;
Middelheimlaan 1, B-2020 Antwerpen, Belgium 
\and
Royal Observatory of Belgium;
Ringlaan 3, B-1180 Brussels, Belgium 
\and
Centre de Recherche en astronomie, astrophysique et g\'eophysique. Route de l'Observatoire BP 63 Bouzareah - DZ-16340 Algiers, Algeria
\and
Institut d'Astrophysique et de G\'eophysique, Universit\'e de Li\`ege; 
All\'ee du 6 Ao\^ut, 17, B-4000 Li\`ege, Belgium
\and
GEPI, Observatoire de Paris, CNRS, Universit\'e Paris Diderot; 
Place Jules Janssen 92195 Meudon Cedex, France 
             }
\abstract 
{The derivation of radial velocities from large numbers of spectra that 
typically result from survey work, requires automation. However, except for 
the classical cases of slowly rotating late-type spectra, existing methods of 
measuring Doppler shifts require fine-tuning to avoid a loss of accuracy 
due to the idiosyncrasies of individual spectra. The radial velocity 
spectrometer (RVS) on the Gaia mission, which will start operating very 
soon, prompted a new attempt 
at creating a measurement pipeline to handle a wide variety of spectral types.} 
{The present paper describes the theoretical background on which this software  
is based. However, apart from the assumption that only synthetic templates are 
used, we do not rely on any of the characteristics of this instrument, so 
our results should be relevant for most telescope-detector combinations.}
{We propose an approach based on the simultaneous use of several 
alternative measurement methods, each  having its own merits and 
drawbacks, and conveying the spectral information in a different way, leading 
to different values for the measurement. A comparison or a combination 
of the various results either leads to a ``best estimate'' or indicates to 
the user that the observed spectrum is problematic and should be analysed 
manually. }
{We selected three methods and analysed 
the relationships and differences between them from a unified point of view; 
with each method an appropriate estimator for the individual random error 
is chosen. We also develop a procedure for tackling the problem of template 
mismatch in a systematic way. Furthermore, we propose several tests for 
studying and comparing the performance of the various methods 
as a function of the atmospheric parameters of the observed objects. 
Finally, we describe a procedure for obtaining a knowledge-based 
combination of the various Doppler-shift measurements.} 
{}

\maketitle

\section{Introduction}

The importance of acquiring the radial velocity (RV) of extra-solar and 
extra-galactic objects is well known. Many techniques that do so, mostly based 
on a measurement of the Doppler shift of their spectra 
\citep[but see e.g.][for an alternative]{DRLM99}, have been established 
during the past five decades, and detailed improvements are still being made. 
In the present era of large spectroscopic surveys such as {\sc rave} 
\citep[e.g.][]{SZSW06,STSZ09} and Gaia \citep[e.g.][]{KMCZ04,WVTM05}, 
the ever-growing amount of available data requires 
increased automation in the scientific analysis, but this may entail  
diminished accuracy or reliability of the final results. The present 
paper revisits the question of Doppler shift measurement with the aim of 
reducing this risk. It provides the theoretical background of the way in 
which RV measurement will be done in 
the {single transit analysis} ({\sc sta}) of the 
Gaia\footnote{http://www.cosmos.esa.int/web/gaia} 
{radial velocity spectrometer} ({\sc rvs}) data, but its 
applicability is not limited to a particular telescope-instrument 
combination. The specific {\sc sta} implementation will be discussed in a 
forthcoming paper. 

\subsection{Historical background}

The idea of using the Doppler effect for determining stellar radial velocities 
dates back to nearly 150 years ago \citep{KLIN66, SOHN67}. The first  
successful attempt to bring it into practice was made by \citet{HUGG68}.  
By the end of the 19$^\mathrm{th}$ century the 
measurement of Doppler shifts was well-established \citep{CORN90} although 
there were still many technical issues to be solved, such as the stability of 
the spectrograph \citep{DESL98}. Until the 1960's one used the position of 
individual spectral lines, as is occasionally still being done for specific 
reasons \citep[e.g.][]{ANNO83,FEKE99}. \citet{FELL55} proposed a 
cross-correlation technique, already well established for analysing 
radar signals \citep[e.g.][]{WODA50,WOOD53}, which would enable the use of all 
the information contained in a spectrum, rather than the position of just 
a few lines. His ideas were developed in practice by \citet{GRIF67} and 
later perfected in dedicated instruments \citep{GRGU74,BAMP79}, all based 
on direct comparison of the spectrum with an optical mask. 

In contrast to these analogue methods, 
\citet{SIMK74} proposed a {digital} cross-correlation;  
in spite of some initial shortcomings, it promised a huge advantage 
in flexibility, error control, and in particular efficiency, owing to 
the use of the {fast Fourier transform} (FFT) for calculating 
the cross-correlation function (CCF). 
This proposal was followed up by several reformulations or rederivations to 
adapt the generic technique for application to data sets with specific 
characteristics \citep[e.g.][]{DAFR77,GRHU77,SSBS77}. Later  
\citet{TODA79} provided an alternative justification based on a $\chi^2$ 
minimization, thus removing some apparent limitations of the method.  
Cross-correlation using FFT eventually became the main ``off-the-shelf'' 
method for RV measurement in many data reduction packages 
\citep[e.g.][]{HILL82,KUMI98}, reaching precisions better than $100\ms$ for 
G- and K-type stars. We refer to it hereafter as the {standard 
method}. Its last significant reformulation is from \citet{ZUCK03}, who 
derives it from a maximum likelihood argument that enables him to provide 
a consistent estimate for the actual random error in a given measurement. He 
also shows how to combine the information from different orders of a 
spectrum without merging them. 
Although it was originally conceived for galaxies and single stars, the 
standard method has also been applied to spectroscopic binaries 
\citep[e.g.][]{HILL93,RAMM04}. A formal extension to deal with double or 
triple spectra was elaborated by \citet{MAZU92}, \citet{ZUMA94} and 
\citet{MZGL95}, and for quadruple spectra by \citet{TOLS07}. 

While the concept and the implementation of the standard method are fairly 
straightforward, its proper use in practice is less so. In fact it requires 
considerable fine-tuning to yield the best possible information, depending 
on the nature of the spectra involved. Late-type single-star spectra are not 
very demanding in this respect, but for early types fine-tuning may have 
to be different from one temperature class to the next in order to reduce 
spectrum mismatch errors, especially if the object is rotating fast 
\citep[e.g.][]{VDGR99,GRDV00}. Of course the availability nowadays of 
high-quality synthetic spectra to serve as templates alleviates this problem 
but even so, spectrum mismatch cannot be entirely excluded and may, in the 
absence of a large number of sharp lines, cause significant errors on the 
measured Doppler shift. 

Several (likewise digital) alternatives for the standard method have been 
proposed; they could be classified roughly as {minimum-distance} 
\citep[e.g.][]{WEJW78,DEMR87,ZWMS05}, {mask} \citep{FUFU90,BQMA96}, 
{phase-shift} \citep{CHEL00} or 
{Pearson-correlation-coefficient} \citep{ROYE99,ZZMI07} methods. \citet{FUFU90} 
showed that the minimization of the sum-of-squares distance proposed by 
\citet{WEJW78} is actually equivalent to the standard method, but the other 
alternatives are truly different (see also Sect.\,\ref{sec:rel}).  

Recent developments were (and are being) driven by the quest for 
exoplanets and by asteroseismology, both of which were envisaged already 
in the 1990's but really took off around the turn of the millennium. These 
studies require measurement precisions better than $10\ms$, which is close 
to the photon-noise limited precision \citep{BMMD96, BOPQ01} even for 
high-resolution spectra of bright slowly rotating late-type stars. Accuracy 
is not an issue here because one needs to detect only a variation of the 
radial velocity, which led \citet{CONN85} to coin the term {accelerometry} 
for this kind of work. Dedicated instruments were built with a focus on 
stability and on precise wavelength calibration using Th-Ar 
comparison spectra \citep{BQMA96,LOPE07}, absorption cells \citep[e.g.][and 
references therein]{BMMD96,KASI08}, and (most recently) laser frequency combs 
\citep[e.g.][]{MUHS07,SWAH08,LBFG08,CFSD10}, which could improve the 
precision of wavelength calibration to the order of $1\cms$. 

The exoplanet search has also shifted observers' attention towards less 
massive stars (because 
their reflex motion must have a larger amplitude), which are best observed 
in the infrared \citep[e.g.][]{SEKA08,MAGE09,RBHD10,VEMM10}. Besides ``plain'' 
spectroscopy, new techniques based on a combination of interferometry and 
spectroscopy have emerged \citep[e.g.][and references therein]{BHCM09,VAGM10}. 
In most cases basically the standard method is still being used for the actual 
Doppler-shift measurement.  

\subsection{Measurement algorithms}\label{sec:alg}

The focus of the present paper is not, however, on such ultra-high precision 
work but on an approach that can yield reasonable (if not truly optimal) 
results for a wide range of spectral types without any object-dependent tuning 
so that it can be fully automated.  
This approach involves building an environment in which several methods 
are applied in parallel and the final outcome is a knowledge-based 
combination of their various results, the ``knowledge'' consisting of 
performance-test results obtained from simulated observations and stored 
in an auxiliary data base. 

The fundamental assumption of any Doppler-shift measurement is that, for a 
given observed spectrum $f$ one has a collection of template spectra $t$ 
that are identical apart from a known Doppler shift and that differ from 
$f$ only by the object's RV and by a random (observational) error on the flux 
samples. Furthermore one assumes that there is some measure 
of similarity or dissimilarity between the spectra, to be maximized, resp. 
minimized over the collection of templates. If no specific choice is 
intended we shall commonly call this measure a {C-function}. In general 
none of the available templates has exactly the same Doppler shift as the 
object, so an interpolation is required to obtain this velocity from the 
C-function samples; this interpolation is often termed {centroiding}. 
Altogether an algorithm for Doppler-shift measurement thus essentially 
consists of a (dis)similarity measure and a method of centroiding. 

The main motivation for implementing more than one algorithm is as  
follows. All C-functions by definition reach their extremum at the same 
template velocity if the latter can be varied continuously and if object and 
template then match exactly (no noise, no spectrum mismatch etc.). However, 
they all respond differently to any deviation from this ideal situation,  
i.e. they differ in their sensitivity to the inevitable error sources (both 
random and systematic) and, if one has to deal with a large variety of 
spectra (as e.g. in the case of Gaia's {\sc sta}) one cannot expect 
any single method to be optimal for all of these. Furthermore, any significant 
difference (i.e. in excess of the expected random error) between the results 
they produce, may provide a useful indication of special care being required 
for the object at hand. Another possible advantage of using alternatives 
to the standard method, is that the latter requires rebinning the spectra to 
equidistant points in ln(wavelength), which may be undesirable if one wishes 
to exploit accurate knowledge of the random error distribution for 
individual observed flux samples (as, again, in the case of Gaia). 

In Sect.\,\ref{sec:cfun} we select three C-functions and we review their 
foundation in a way that highlights the relations and differences 
between them. Sect.\,\ref{sec:meas} 
discusses {centroiding} and the so-called {model mismatch error} 
it may entail. Sect.\,\ref{sec:tmerr} deals with 
the problem of {template mismatch}. For each of the measurement 
algorithms an appropriate random-error estimate is chosen 
in Sect.\,\ref{sec:rerr}.  In 
Sect.\,\ref{sec:tests} we discuss several tests for assessing the 
performance of the algorithms, as well as the storage of the test 
results. In Sect.\,\ref{sec:comb} we use the latter to combine the various 
measurements resulting from the chosen methods.  Finally, Sect.\,\ref{sec:con} offers some conclusions.  

Note that we do not consider any corrections for the gravitational 
redshift, convective motions etc., so the term {radial velocity} 
used here actually refers to the {``barycentric radial velocity 
measure}'' defined by \citet{LIDR03}, 
not to the true line-of-sight velocity as determined from e.g. astrometric 
observations \citep{DRLM99,LIMD00}. Such corrections, if required 
scientifically, will be assumed to be built into the library of synthetic 
templates one uses or to be added outside the framework of the Doppler-shift 
measurement.

\section{Foundation of the C-functions}\label{sec:cfun}

The C-functions we consider for a multi-method approach are the standard CCF 
(Sect.\,\ref{sec:cc}) because its computation using FFT is much faster than 
the others, the Pearson correlation function proposed by \citet{ROYE99} 
(Sect.\,\ref{sec:pc}) because it is very flexible, not requiring the observed 
fluxes to match the template in an absolute sense, and a function derived from 
the maximum likelihood principle (Sect.\,\ref{sec:ml}). 
We chose to concentrate on algorithms that are applicable to as wide a 
variety of spectra as possible, so we do not consider a phase-shift method 
because its applicability would be limited to very high $S/N$, nor a 
digital-mask method because it would require high resolution. Nevertheless, 
these algorithms do have their merits and, if it is judged useful for some 
reason, they could easily be included in a multi-method approach.

\subsection{Notations and definitions}\label{sec:def}

Throughout this paper we use $\lambda$ to represent wavelength in general, 
$\lambda_n$ for the central wavelength of the $n^\mathrm{th}$ bin of an 
observational grid, $\Delta_n$ for half the width of this bin (in general 
the width will vary along the spectrum), $f(\lambda)$ for the spectrum 
emitted by an observed object, $f_n$ for its $n^\mathrm{th}$ flux sample 
(which is of course the only real information we 
have about this spectrum), $\sigma_n$ 
for the random-error estimate on this sample, $t(\lambda;v)$ for a template 
spectrum (see Section~\ref{sec:prop2}) that has been calculated for a source 
moving with radial velocity $v$ (called the {template 
velocity}), and $t_n(v)$ for the $n^\mathrm{th}$ sample of the latter, binned 
to the observed wavelength grid, i.e.  
\beq
t_n(v)=\int_{\lambda_n-\Delta_n}^{\lambda_n+\Delta_n}
t(\lambda;v)\dff\lambda  \ .                     \label{eq:giv}
\eeq 
Then, assuming that the template exactly matches the source at hand, 
we can state that the observed fluxes are related to the template by 
\beq
f_n = a_\mathrm{s}t_n(v_\mathrm{s}) + d_n    \ ,                \label{eq:rfg}
\eeq
where $a_\mathrm{s}$ is merely a scale factor (e.g. for the conversion of 
units or to allow for some uncertainty in the continuum level of the object) 
and $v_\mathrm{s}$ is the radial 
velocity of the source, both to be determined from the data, while $d_n$ is 
a random variable with zero mean, representing the noise in the 
$n^\mathrm{th}$ observed sample. If the scale factor $a_\mathrm{s}$ is known 
a priori one can include it in the definition of 
the template and put $a_\mathrm{s}=1$ in the expressions to be derived below. 

The term {data segment} may refer to a set 
of observed fluxes $(f_1,f_2,\dots,f_N)$ for a given object as well as to 
a set of ``predicted'' fluxes $(t_1(v),t_2(v),\dots,t_N(v))$. 

If a data segment has been rectified so that its (pseudo) continuum level 
is constant, and if its ends lie within a line-free region of the 
spectrum \citep[see][for more details]{VEDA99}, it will be termed 
{regular}. Even though in practice we cannot expect all spectra to be 
truly regular, this notion will be useful in the discussion below.    

\subsection{Properties of the input data}
\subsubsection{Observed flux}\label{sec:prop1}

We assume that 
\begin{enumerate}
\item{} the observed spectra have been prepared 
(subtraction of bias, dark current, and celestial background; detection of 
cosmic-ray effects and CCD blemishes; extraction; wavelength calibration)  
in both normalized and non-normalized form, the latter implying that the 
value of each sample is expressed as a number of photons (but note 
that this number is generally non-integer after the data preparation); 
\item{} the random errors on the flux samples are statistically independent; 
\item{} the probability density function (PDF) of this random error is 
Gaussian with known ``real'' dispersion for each sample, obtained 
e.g. from standard error propagation taking account of all relevant noise 
sources (photon count, read-out, bias, dark current, background etc.); 
however, to provide for a situation where such error estimates 
are not available we also discuss two alternative models for this PDF 
(see Sect.\,\ref{sec:ml});
\item{} flux samples affected by cosmic-ray hits or CCD blemishes have 
been given some reasonable value but they are also marked so that (where 
necessary) they can be excluded from the calculations; 
\item{} any wavelength regions containing spectral features that originate 
from matter that does not share the motion of the source (e.g. interstellar 
lines) have been properly identified and the flux samples within them are 
marked as above; 
\item{} the spectra are not noise-filtered because that would inevitably 
destroy the assumed statistical independence of the flux samples.  
\end{enumerate}
We do not presume that the spectra are contiguous: one may determine  
separate C-functions for e.g. the orders of an echelle spectrogram, 
different CCDs (as in the case of Gaia) or disjoint wavelength regions 
(e.g. to eliminate interstellar or atmospheric spectral features); 
afterwards these C-functions can be averaged \citep{ZUCK03} or, for 
those of Sect.\,\ref{sec:ml}, summed.

\subsubsection{Template}\label{sec:prop2}
In principle we assume that 
\begin{enumerate}
\item{} the atmospheric parameters (APs) of the observed object are known 
and a corresponding synthetic spectrum $s(\lambda)$ is available with 
virtually infinite resolution; the former condition is seldom realistic so in 
Sect.\,\ref{sec:tmerr} we discuss the consequence of relaxing it; 
\item{} the projected rotational velocity of the object is known and the 
synthetic spectrum has been broadened accordingly; 
\item{} the template spectrum $t(\lambda,v)$ is obtained by applying the 
Doppler shift, smooth\footnote{i.e. insofar as it can be described by a 
factor that is constant or at most slowly varying over the wavelength 
range considered.  }
interstellar extinction and all known instrumental effects (optical 
distortion, line-spread function (LSF), CCD 
recording, \dots) to the (rotationally broadened) spectrum $s(\lambda)$; 
it is provided with and without normalization;  
\item{} the template is calculated on a wavelength range that extends 
sufficiently far beyond the observed one so that after any foreseeable 
Doppler shift, the former still completely covers the latter.
\end{enumerate} 

The second assumption is made for the following reason. \citet{TODA79} have 
shown how line-broadening due to velocity dispersion in the spectrum of 
a galaxy may be determined along with the radial velocity in a two-dimensional 
optimization. In principle their approach can be applied also to rotational 
broadening in a stellar spectrum \citep[e.g.][]{DGLG11}. 
However, this may entail some 
correlation between the two measured values, which is often undesirable from 
the point of view of error propagation. Therefore we assume that $v\sin i$ 
is obtained from an independent measurement, as e.g. in 
\citet{RGFG02,GRAY05,CMBP11}.  
The third assumption ensures that Eq.\,(\ref{eq:rfg}) is strictly 
applicable. This may not be required by some of the methods we discuss here, 
but of course the template must be the same for all of them if we want a 
valid comparison between their respective outcomes. 

Formally the template may be represented as  
\beq
t(\lambda;v) = \int_0^\infty G(\lambda,\lambda_\mathrm{s}(\Dsh))
s(\lambda_\mathrm{s})\dff\lambda_\mathrm{s} \ ,              \label{eq:tmpl}
\eeq
where $G$ is a Green's function, $\lambda$ is the wavelength in the 
observer's rest frame and $\lambda_\mathrm{s}$ is the wavelength in the 
rest frame of the synthetic spectrum. 

This form implies that the integration should be performed for every 
value of $v$ 
considered. However, in many cases this integral is well approximated by a 
simple convolution of $s$ with the LSF, multiplied by a scale factor. 
If the LSF is sufficiently narrow one can show that actually 
\beq
t(\lambda;v) \approx \frac{1}{\Dsh}t(\frac{\lambda}{\Dsh};0)  \label{eq:giv0}           
\eeq
so that the convolution must be computed only once and Eq.\,(\ref{eq:giv}) 
can be written as 
\beq
t_n(v) = 
\int_{\frac{\lambda_n-\Delta_n}{\Dsh}}^{\frac{\lambda_n+\Delta_n}{\Dsh}}
t(\lambda;0)\dff\lambda  \ .                                    \label{eq:giv1}
\eeq
In situations where Eq.\,(\ref{eq:giv0}) is no longer valid for large 
template velocities, e.g. due to charge distortion effects caused by 
radiation damage (which cannot be accounted 
for in a convolution), one can compute Eq.\,(\ref{eq:tmpl}) on a coarse 
velocity grid and apply an approximation such as Eq.\,(\ref{eq:giv0}) only to 
deviations from these grid values.

\subsection{C-functions}\label{sec:CFS}

\subsubsection{Cross-correlation -- the standard method}\label{sec:cc}

The {cross-correlation function} (CCF) is a descriptive 
tool developed for investigating stochastic processes. 
In her pioneering paper on the digital measurement of Doppler shifts, 
\citet{SIMK74} referred to the standard literature in this field
\citep[e.g.][]{JEWA69,BEPI71} to introduce the use of cross-correlation for 
Doppler-shift measurements, stressing the necessity of rectifying 
\citep{JEWA69} the spectra and of rebinning them to a grid with constant 
step in $\ln\lambda$ (see Appendix~\ref{sec:tmpls}) so that they can be 
considered as a so-called {stationary random process}.    

In our notation, taking account of 
Eq.\,(\ref{eq:dsh}), the CCF as defined by \citet{TODA79} is 
\beq
C_\mathrm{cc}(m) = \frac{\sum_{n=1}^N (f_n-\bar{f})(t_{n-m}(0)-\bar{t}(0))}
{\sqrt{\sum (f_n-\bar{f})^2\sum (t_{n-m}(0)-\bar{t}(0))^2}}    \ ,  
                      \label{eq:CCF}
\eeq
where $\bar{\ }$ indicates the average over a data segment. Note that 
$f_n$ and $t_n$ are considered here as periodical functions of $n$ with 
period $N$; this implies that indices $n-m<1$ or $n-m>N$ correspond to 
{wrapped-around} values of $t_n$ with $1\leq n\leq N$. This 
wrap-around may cause a systematic error \citep{SIMK74} unless both 
data segments are regular and the shift does not exceed the length of the 
line-free region at either end. In case of larger shifts one may therefore 
need to make a crude measurement first and then use an adapted data segment 
for the template to do the final measurement. 

Although we adopt this definition, for stellar spectra we do not follow 
most of the data preparation that \citet{TODA79} advocate: 
\begin{itemize}
\item{}instead of continuum subtraction we prefer normalization, which 
serves the same purpose while conserving the relative strength of the lines;
\item{}both high- and low-frequency filtering are omitted;
\item{}we prefer to avoid endmasking (or apodization) because this 
introduces a structure in the observed 
spectrum that does not share the Doppler shift of the stellar features, 
thereby possibly causing a systematic error \citep[p.\,131]{Vers91}; the 
purpose of endmasking is better served by adapting the wavelength range 
(if possible) to make the data segments regular. 
\end{itemize}

In the standard method one does not evaluate the sum in Eq.\,(\ref{eq:CCF}) 
directly, but via its Fourier transform, using the so-called 
Fast-Fourier technique; in that case the number of flux samples in each 
data segment must be a power of two. This may be achieved by adding, at 
either end, a number of samples having the continuum value (an operation 
also known as ``zero-padding'') or by adapting the stepsize in 
$\ln\lambda$, provided that the adopted stepsize 
can be kept close to the average observed bin width.  

If the observed spectrum contains samples marked as invalid (see 
Sect.\,\ref{sec:prop1}), one has two options: either to use some 
reasonable replacement value for the invalid samples or to split up 
the data segment into smaller ones that do not contain any ``bad'' data. 
The latter option is preferable if several adjacent pixels are 
affected but it is hardly practicable in an automated treatment.    

\subsubsection{Pearson correlation}\label{sec:pc}

\citet{ROYE99}, while referring to \citet{SIMK74} and \citet{TODA79}, 
actually adopted a quite different approach using the classical Pearson 
correlation coefficient as a measure of the similarity of the data 
segments representing respectively the object and the template 
calculated for a particular template velocity $v$. This implies that 
he considered these data segments merely as ordered sets of measurements 
of two correlated random variables or, rearranged in the form 
$\left\{(f_n,t_n(v))|n=1,2,\dots N\right\}$, as an ordered set of values 
drawn from a bivariate population with a certain correlation. The 
distribution of this population would be nondescript\footnote{Incidentally, 
it would look Gaussian-like in most parts of late-type spectra, but 
otherwise its shape could be very different.}  
but we note that anyway all its moments would be formally finite. If 
the observed 
spectrum is also noiseless and different only by the radial velocity of 
its source, the correlation will depend only on the object--template 
velocity difference and on the wavelength binning, becoming a strict 
equality if the velocities are equal.  

\citet{ROYE99} prepared his data with the logarithmic sampling 
Eq.\,(\ref{eq:clnl}), but this is not required in principle. In fact we can 
define the resulting C-function in general as 
\beq
C_\mathrm{pc}(v) = \frac{\sum_{n=1}^N (f_n-\bar{f})(t_n(v)-\bar{t}(v))}
{\sqrt{\sum_{n=1}^N (f_n-\bar{f})^2\sum_{n=1}^N (t_n(v)-\bar{t}(v))^2}} \ ,
                                                              \label{eq:PCF}
\eeq
This 
expression presumes nothing about the wavelength bins in Eq.\,(\ref{eq:giv}), 
except that the grid $(\lambda_1,\lambda_2,\dots ,\lambda_N)$ is common 
to both spectra. Sample indices $n$ that have been marked as invalid in 
the observed spectrum (see Sect.\,\ref{sec:prop1}) are simply to be omitted 
from all sums (also from those containing only template data).    
Hereafter Eq.\,(\ref{eq:PCF}) will be termed the 
{Pearson correlation function} (PCF) as distinct from the CCF. 

Unlike the CCF, the PCF does not require that the data segments are regular 
or even rectified\footnote{This does not imply that, for a given observed 
spectrum, the Doppler-shift value measured on the normalized form should be 
{strictly identical} to the one obtained with the non-normalized form, 
but the difference is usually much smaller than the common measurement errors.} 
so it should be applicable also where e.g. rectification 
is problematic. Nevertheless one should be aware that its sensitivity is 
likely to be degraded if the ratio of the continuum levels of object and 
template varies considerably over the spectral range. 

We also note that, in its general form with an arbitrary wavelength 
grid for the observed data, the PCF can be evaluated only in  
velocity space; this is in fact the case we consider hereafter.

\subsubsection{Maximum likelihood / minimum distance}\label{sec:ml}

The noise in different samples is uncorrelated but of similar
origin so we can consider a data segment as a measurement of a set of 
$N$ independent random variables that are distributed similarly, their  
frequency functions differing only by their expectation $e_n$ and dispersion 
$\sigma_n$ where, according to Eq.\,(\ref{eq:rfg}),     
\beq
e_n=a_\mathrm{s}t_n(v_\mathrm{s})\ \ ,\ \ \  
\sigma_n=\mathrm{stdev}(d_n) \ .                           \label{eq:ed}
\eeq 
We note $\mathrm{pd}(f_n|a,v)$ for their individual PDF.  
Since the form of this function is 
known or can be reasonably hypothesized, it is justified to use the maximum 
likelihood (ML) principle for estimating the parameters $a_\mathrm{s}$ 
and $v_\mathrm{s}$ \citep{ZUCK03}. Given a data segment with $N$ samples, 
the likelihood function is given by 
\beq
L(a,v|f) = \prod_{n=1}^N\mathrm{pd}(f_n|a,v) \ .             \label{eq:llh}
\eeq
Invalid sample indices should be omitted from this product (and thus from 
all sums derived from it hereafter).  
The parameter values $(\hat{a},\hat{v})$ that maximize this likelihood 
constitute a consistent asymptotically minimum-variance estimator for the 
set $(a_\mathrm{s},v_\mathrm{s})$ 
\citep[see e.g.][and references therein]{MART71}.  

Whereas the PCF essentially measures the similarity of the flux gradients, 
the likelihood function measures the similarity of the fluxes themselves. 
Therefore a method based on the ML principle is also a {minimum 
distance} method, which is a more significant name in the present context. 

\citet{ZUCK03} assumes exclusively that the distribution is Gaussian and 
that the dispersions $\sigma_n$ are all equal but unknown; he determines 
this unknown value $\sigma_\mathrm{s}$ from the data along with the other 
parameters. However, in Sect.\,\ref{sec:prop1} we made other assumptions so  
we need to derive the appropriate equations for our case. If the PDF is 
Gaussian with known dispersion $\sigma_n$ for each sample, then 
\beq
\mathrm{pd}(f_n|a,v) = \frac{1}{\sqrt{2\pi}\sigma_n}
  e^{-\frac{(f_n-at_n(v))^2}{2\sigma_n^2}}   \ .           \label{eq:gpdf}
\eeq
Taking the logarithm of Eq.\,(\ref{eq:llh}) one easily sees that maximizing 
the likelihood is equivalent to minimizing  
\beq
C_\mathrm{md}(a,v) = \sum_{n=1}^N \frac{(f_n-at_n(v))^2}{\sigma_n^2}
                                                           \label{eq:gdis}
\eeq
which, incidentally,  equals the classical $\chi^2$ statistic for a 
goodness-of-fit test. Obviously this expression can in fact be seen as an 
error-weighted squared ``distance'' between object and template. 

As pointed out before, if $a$ is known independently from the minimization 
problem, it may be omitted from this expression. Otherwise the 
two-dimensional minimization to determine $(\hat{a},\hat{v})$, can be 
reduced to one dimension by using the fact that the derivative 
$\pderI{C_\mathrm{md}}{a}$ must vanish at the minimum; this leads to    
\beq
\hat{a} = \frac{\sum\frac{f_nt_n(\hat{v})}{\sigma_n^2}}
{\sum\frac{t_n^2(\hat{v})}{\sigma_n^2}}      \ .               \label{eq:has}
\eeq
Substituting this in Eq.\,(\ref{eq:gdis}) we find that $\hat{v}$ is the 
value that maximizes 
\beq
C_\mathrm{md1}(v) = \frac{\left|{\sum\frac{f_nt_n(v)}{\sigma_n^2}}\right|}
{\sqrt{\sum\frac{t_n^2(v)}{\sigma_n^2}}} \ .             \label{eq:cv}
\eeq
Having determined $\hat{v}$, one finds $\hat{a}$ from Eq.\,(\ref{eq:has}). 

If, for some reason, the error estimates $\sigma_n$ are not available 
while, on the other hand, it would not be justified to assume that they 
are all equal, one can adopt an alternative model for the flux 
distributions, such as 
\begin{itemize}
\item{} a Gaussian with a predicted dispersion describing photon noise 
(derived from the template) and possibly a stationary additive error source 
(e.g. read-out noise); 
\item{} the Poisson probability function, applicable if one can assume that 
the recorded flux samples represent a pure photon count.  
\end{itemize}
Appropriate minimum-distance equations for these models are given in 
Appendix~\ref{sec:fldist}. 

\subsection{Similarity between different C-functions}\label{sec:rel}

While all the functions defined above are obviously different from each 
other, there is some (at least superficial) similarity between them, and   
it is worthwhile to explore this in more detail. If the binning is 
logarithmic, if $t_n(v)$ is calculated for the discrete 
velocities given by Eq.\,(\ref{eq:vis}) and if both the object and the 
template (for all template velocities considered) are regular as defined 
in Sect.\,\ref{sec:def}, then $\sum {t_n(v)}$ and $\sum {t_n^2(v)}$ are 
constant over a limited velocity range around the object's velocity; 
in that case neither  
the denominator in Eq.\,(\ref{eq:PCF}), nor $\bar{t}(v)$ in the numerator, 
influence the extremum position so that $C_\mathrm{pc}(v)$ becomes 
equivalent to $C_\mathrm{cc}(v)$ (i.e. both must yield {exactly} 
the same value for $\hat{v}$). 
Likewise for such a dataset, if moreover all error estimates are equal 
(i.e. $\sigma_n = \sigma_0\ \forall\, n$) then also $C_\mathrm{md1}(v)$ 
becomes equivalent to the former two. 

With any new implementation of these (and possibly other) methods it is 
advisable to test the above, first with 
noiseless data so that the results must agree within the known bounds of 
machine precision, then with some noise added to obtain a first indication 
of how each method reacts to that.  If these tests are satisfactory one can 
consider gradually more realistic cases and so acquire a good understanding 
of the differences between the measurements and of their relation 
to the characteristics of the observed spectra (atmospheric parameters, 
rotational broadening, resolution, $S/N$-level, \dots).

\section{Measurement}\label{sec:meas}
\subsection{Identifying the peak}

In many applications the range of velocities one expects to encounter will be  
very wide; nevertheless on physical grounds it must be possible to estimate  
reasonable boundaries for this range, say, $[v_\mathrm{min},v_\mathrm{max}]$. 
Then a first (coarse) estimate of the Doppler shift can be made as follows. 

With the methods operating in velocity space one calculates the C-function 
over the whole velocity range $[v_\mathrm{min},v_\mathrm{max}]$ on 
a relatively coarse grid with step size sufficiently small to detect 
all local extrema (e.g. $10\kms$ for spectra with instrumental resolution 
11500) and one locates the 
highest/lowest value. Then, reducing the range to one coarse step to either 
side of the latter and the step size with some factor (e.g. 10), the location 
is improved. If necessary a further refinement may be obtained in the same 
way. The actual measurement is done by centroiding the extremum within the 
resulting (small) set of C-function samples.    

On the other hand, the standard method implies that a {fixed} 
segment of the template spectrum is ``shifted'' along the observed one. 
As a consequence, if the source has very large RV, the template 
segment cannot match the observed spectrum very well (even at the correct 
shift) because either will contain a part of the spectrum that the other 
does not. 
Moreover the CCF is not truly refined by choosing a smaller velocity step 
(i.e. a smaller bin width in $\ln\lambda$) because that would require 
resampling also the observed spectrum to smaller bins, which may imply an 
interpolation of mere noise fluctuations. For these reasons we proceed as 
follows.   

First consider the two ``extreme'' template data-segments, obtained with  
$v_\mathrm{min}$ and $v_\mathrm{max}$ resp. as the template velocity. 
Cross-correlating each of these with the observed spectrum should yield \\
- either two shift values that are comparable \\
- or widely different values, one of which corresponds to a much more 
significant peak in the CCF than the other (e.g. difference in height 
exceeds five times the expected random error on the CCF values). \\ 
If (in an automated treatment) neither of these situations occur 
then manual intervention is advisable. With the above prescription 
one can obtain an indication of whether the RV is 
much smaller than the width of the whole interval, or comparable to it; in 
the latter case it is best to use a template data segment calculated on 
a $\ln\lambda$ interval that has been pre-shifted according to the coarse 
estimate, so that the residual shift to be found by centroiding is relatively 
small. 

It should be noted here that with very noisy spectra  
the highest/lowest C-function value identified as ``the peak'' may be totally 
unrelated to the Doppler shift so that the above procedures (and indeed the 
measurement as such) become meaningless. This situation, where the observed 
extremum position may deviate much more from the true Doppler shift than 
predicted by any error estimate, is discussed in Sect.\,\ref{sec:vfo}.

\subsection{Centroiding}\label{sec:cen}

Centroiding is essentially the interpolation between a number of discretely 
sampled C-function values with a view to find the position of the extremum 
of the underlying (continuous) function. \citet[][section 5.7.2]{RAMM04} and 
\citet{KUMI98} list a number of interpolating functions that can be used for 
this purpose. However, any interpolating function that does not equal 
the underlying function (up to a shift) may itself be a source of 
error, the so-called model-mismatch error \citep[MME,][]{DAVE95}, 
and generally makes the result sensitive to the number of samples (or 
{``fit points''}) used for determining the parameters of the 
interpolating function. 

As in any interpolation problem the accuracy of the result improves if the 
interpolating function better resembles the underlying function. Therefore 
any information one has on the intrinsic nature of the C-function, may be 
useful in choosing an appropriate model. For instance if one knows that the 
C-function is intrinsically symmetric with respect to its extremum, obviously 
the interpolating function should have the same property. And if moreover 
one has reasons to expect that the C-function is close to e.g. Gaussian (as  
with slowly rotating G stars) one may try that as a model. Nevertheless it 
has been noted \citep{ALLE07} that even here the Gaussian (which in 
principle is more robust against noise) does not always perform better than 
a simple parabola. This is related to the fact that a Gaussian requires 
fit points down to the base level, which do not contain any information on 
the Doppler shift.  
 
The number of fit points to be used, depends on several circumstantial factors. 
First of all, it is obvious that ``clean'' information on the Doppler shift 
resides only very close to the centre of the C-function peak or trough, and 
certainly not farther away from it than the HWHM of the sharpest lines 
in the observed spectrum.  
 
If the C-function sampling step is much smaller than this maximum distance, 
one may be tempted to use as many fit points as possible, to 
suppress the effect of random errors, but 
this makes sense {only} if the random error on adjacent C-function 
values is statistically independent; that may be the case, approximately, 
in the standard method where the C-function sampling is tied to the
observed wavelength bin, but it can hardly be true with the velocity-space
methods, which refine the velocity step to a fraction 
of the value corresponding to the observed wavelength-bin width. Moreover, 
if the model does not match the intrinsic (i.e. noise-free) shape of 
the C-function, systematic deviations between those two will gain 
influence when fit points farther away from the extremum are used. 

As a general rule, assuming we do not know the exact shape of the 
C-function, we thus find that it is best to use the simplest possible model 
(i.e. a parabola) and the minimal number of fit points (i.e. 3), unless 
there is clear evidence indicating otherwise; this conclusion fully 
agrees with the results of \citet{ALLE07}. The MME in this case, 
assuming the C-function is locally symmetric, is discussed in 
Appendix~\ref{sec:ubsym}.

One may object to the use of a parabola because this is a symmetric function 
while the velocity-space C-functions are slightly asymmetric so that we 
would knowingly cause a MME. However, the random error on the 
C-function samples generally introduces a much stronger asymmetry, so that 
the use of e.g. a third-degree polynomial would mean that the model will 
mainly pick up noise effects, degrading the actual position measurement 
\citep[cf.][]{ALLE07}. If there is a compelling reason to allow some 
asymmetry in the model, one should use at least a polynomial of even degree 
\citep{VEDA99}. 

On the other hand, knowing that the use of a parabola for centroiding an 
asymmetric C-function may thus cause a somewhat larger MME than 
the basic one discussed in Appendix~\ref{sec:ubsym}, we need a way to check 
whether it is in fact still negligible in a given situation. 
Therefore in Appendix~\ref{sec:ubasy} we derive (under fairly mild 
conditions) an upper bound for this error.   

\section{Template mismatch error }\label{sec:tmerr}
\subsection{General remarks} \label{sec:tmgen}
The template used for measuring the RV of a given single star, is derived 
from a synthetic spectrum, selected from a library using the object's 
atmospheric parameters and rotationally broadened with an estimate of 
the object's $v\sin i$. 

Under most circumstances, even if some of the APs could be determined with 
very high accuracy, their value will not correspond {exactly} 
to any of the available synthetic spectra. Given the almost inevitable lack 
of knowledge on several other factors determining the details of an 
observed spectrum, it seems pointless to consider interpolating between  
library spectra. Therefore we assume that simply the one nearest (in AP space)  
to the observed one, will be used. This means that in principle each of 
the chosen template's APs could be in error by 1/2 its step 
size in the library, even under otherwise ideal circumstances. 

Almost any mismatch between template and object may bias the RV measurement. 
Such bias is usually referred to as the {template mismatch error} 
(TME). In principle this is a systematic error, since it will have the 
same sign and order of magnitude for all observed objects 
with the same intrinsic AP values. Nevertheless, the uncertainty on the APs 
being random, the user of the RV measurements may decide to treat the TME 
as if it were an additional random error.  

To estimate the TME, \citet{NMKD02} measure each object against a {
set of templates} rather than just the best matching one; the resulting 
set of RV-values will then provide an estimate of the TME. This would 
surely suffice if one has to deal with only a small number of objects, but 
e.g. in survey work one may prefer a quick estimate obtained from a table or 
from a simple model, even though inevitably any such estimate would be less 
realistic.   

Therefore in this section we propose a strategy for studying, with a given 
synthetic library and in a given region of AP space, how the TME may behave 
both as a function of the AP errors (later referred to as a {local 
model}) and as a function of the spectral type of the object (a {global 
model}). This information should help a user of the RV measurements to 
assess the importance of the TME in view of the accuracy requirements and 
of the random uncertainty of the RV. 

In the examples below we consider only the effect of an uncertainty in three 
APs {and in $v\sin i$, but it is clear that other sources of mismatch, 
such as the convective deformation of line profiles \citep[see e.g.][and 
references therein]{GULI02} could be treated in a similar way. 

\subsection{Modelling strategy} \label{sec:strat}
\subsubsection{The AP grid }\label{sec:grid}
The dominant APs determining the spectral features 
are the temperature ($T_\mathrm{eff}$), gravity ($\log g$), and 
metallicity ([Fe/H]). Other APs such as [$\alpha$/Fe], turbulent velocity, 
the abundance of specific elements or molecules, 
magnetic field strengths, etc. may be relevant in particular cases, 
depending on the required accuracy, but in the present global 
assessment they will be ignored. Thus we limit ourselves to a 3-dimensional 
AP-subspace within which the triplets ($T_\mathrm{eff}$,\,$\log g$,\,[Fe/H]) 
characterizing the available library spectra, define a grid with (generally) 
variable stepsize. 

Unlike the APs, the $v\sin i$-value to be applied is not 
{technically} limited to a 
discrete set; however, this value inevitably has some measurement error 
and we need to investigate how that may affect the final RV-error.   
Therefore, in the discussion below we shall treat $v\sin i$ on an equal 
footing with the atmospheric parameters, considering its estimated value 
as situated on a grid with local stepsize equal to twice the uncertainty of 
this value. As a consequence, formally we shall consider a 4-dimensional 
grid with each node characterized by a set of values 
$c=(T_\mathrm{eff}$,\,$\log g$,\,[Fe/H],\,$v\sin i )$. 

In physical terms this grid is not uniform, but we shall 
use it mainly for bookkeeping, so whatever the actual amount 
that is represented by a given edge, 
we consider the latter to have length ``1'' in the appropriate units. 
Assuming for the sake of simplicity that the AP steps are locally constant, 
we arrange these in a local ``units'' vector $u=(u_1,u_2,u_3,u_4)$, e.g. 
\beq
u=(250\,\mathrm{K}, 0.5\,\mathrm{dex}, 0.5\,\mathrm{dex}, 
10\kms)    \ .                    \label{eq:apdev}
\eeq
Noting $(c_1,c_2,c_3,c_4)$ for the node chosen as the origin, any 
set of AP's in the latter's vicinity can be characterized by a vector 
$x=(x_1,x_2,x_3,x_4),\ x_i\in\RR$, the actual parameter values being 
given by $a_k = c_k + x_ku_k \ , \ \ \ k=1,\dots 4$. 
With these definitions we can think of the grid as a simple hypercubic
lattice.  

\subsubsection{Simulating the TME }\label{sec:e_tme}
We choose a given node in the AP grid as its origin and consider the 
corresponding template spectrum, prepared as in Sect.\,\ref{sec:prop2} at 
zero velocity, as an ``observed'' spectrum. Then we measure the 
latter's Doppler shift against the templates corresponding to the 
neighbouring nodes; all the spectra involved should be noiseless, since 
we are studying a systematic effect here. The resulting values constitute 
a local (discrete) sampling of the TME as a function of AP deviations 
$\Delta a_k=x_ku_k \ , \ \ \ k=1,\dots 4$. Fig.\,\ref{fig:1000K} shows 
a simple example (only $T_\mathrm{eff}$ is variable) of such a function; 
notice that the 
effect of deviations with $|x_1|\leq 2$ could be described by a 
$2^\mathrm{nd}$-degree polynomial in  $x_1$ but that larger deviations 
would require a $3^\mathrm{rd}$-degree polynomial.   
\begin{figure}
\includegraphics[width=\columnwidth]{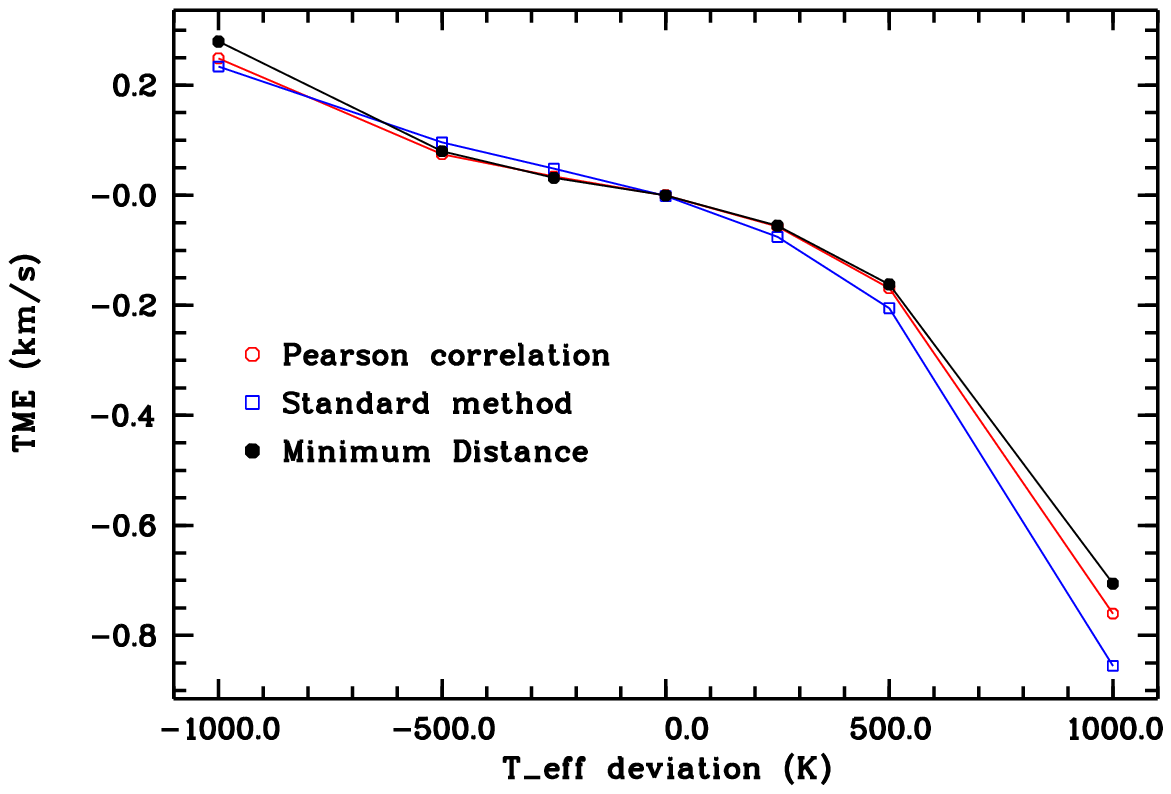}
\caption{Simulated TME for $c=(5500\,\mathrm{K}, 4\,\mathrm{dex}, 
0\,\mathrm{dex}, 20\kms)$ with a local grid defined by Eq.\,(\ref{eq:apdev}) 
and $x_2=x_3=x_4=0$ while $x_1=-4,-2,-1,0,1,2,4$.      
}
\label{fig:1000K}
\end{figure}
The following sections discuss the modelling of such functions. 

\subsection{Local model }\label{sec:locmod} 
Lacking any theoretical information about the functional dependence of the 
TME on the AP deviations, we can only assume that it is analytic so that 
the TME may be approximated by a truncated Taylor expansion. 
Approximations of increasing order can be thought of as taking into account 
``cross-talk'' between different APs (i.e. the deviation in one parameter 
modifying the effect of a deviation in another parameter) of increasing 
complexity. E.g. in the first-order the effect of simultaneous 
deviations in several parameters is described simply as the sum of their 
individual effects; in a second-order approximation the combined effect of 
pairwise deviations is described more accurately, etc. 
We shall assume that a second-order approximation, 
\beq
P^{(2)}(x) = \sum_{k=1}^4 C_kx_k + \sum_{k=1}^4\sum_{l=1}^k C_{kl}x_kx_l 
                                          \ ,             \label{eq:2nd}  
\eeq
is sufficient for the 
purpose at hand. As one can see from Fig.\,\ref{fig:1000K}, this assumption 
limits the size of the deviations for which the model is valid. 

The coefficients $C_k$ and $C_{kl}$ in $P^{(2)}(x)$ 
can be obtained from TME measurements for the nodes characterized by 
a vector $x$ that has three or two components equal to zero while the 
remaining ones are $\pm 1$. There are 32 of these for the 14 coefficients, 
but we prefer to keep this redundancy and to obtain the coefficients
from a least-squares fit to the larger set of data, because a matching 
number of nodes would be less symmetrically distributed over the hypercube 
and because in practice some of the nodes may be unavailable anyway owing 
to local incompleteness of the synthetic spectrum library. Actually such 
incompleteness may cause the system to be underdetermined even if nominally 
the number of data exceeds the number of unknowns; therefore in the 
polynomial fit we use a procedure based on the so-called {Moore-Penrose 
pseudo-inverse} \citep[e.g.][]{PENR55,PENR56}, which allows the user to 
make an informed decision on which monomials to remove from the model if 
some coefficients cannot be uniquely determined.  

Once the coefficients for the second-degree model have been obtained, the 
latter can be validated by verifying that, if applied to {all 
adjacent} nodes (of which there are 48 in addition to the 32 used for the 
fit), its residuals do not exceed a pre-set threshold. If the model is 
found to satisfy this requirement it may be trusted to predict the TME 
on an actual RV measurement in which the AP deviation is within the 
range described by the local units vector $u$. 

\subsection{Global model} \label{sec:glomod}
The coefficients in the local model vary with the APs defining the 
central node. This variation should be investigated as well. 
Depending on the outcome of such a study, one may decide either to search 
for a global model or to establish a look-up table for the polynomial 
coefficients. A global model would mean that the 14 coefficients are 
represented, in their turn, by a function of the APs 
$(c_1,c_2,c_3,c_4)$ of the central 
node. However, exploratory calculations (see e.g. Fig.\,\ref{fig:c1}) 
indicate that such a function could not be a low-order polynomial if it 
is to be valid over at least a sizeable region of AP space. Therefore 
we believe it is better to define a relatively coarse grid covering 
all of the relevant AP space, determine the coefficients in Eq.\,(\ref{eq:2nd}) 
for each of its nodes and store these in an auxiliary database. When the 
local model is required for any other node, its coefficients may then be 
approximated by interpolation between those on the coarse grid. Obviously 
such a database will be valid only for a given telescope-instrument 
combination, i.e. for a particular mean instrumental resolution, 
wavelength range and --sampling step. 

Incidentally, Fig.\,\ref{fig:c1} also indicates that the TME may differ 
significantly between measurement methods, which is something one may 
wish to take into account in the combination of their respective results. 

\begin{figure}
\includegraphics[width=\columnwidth]{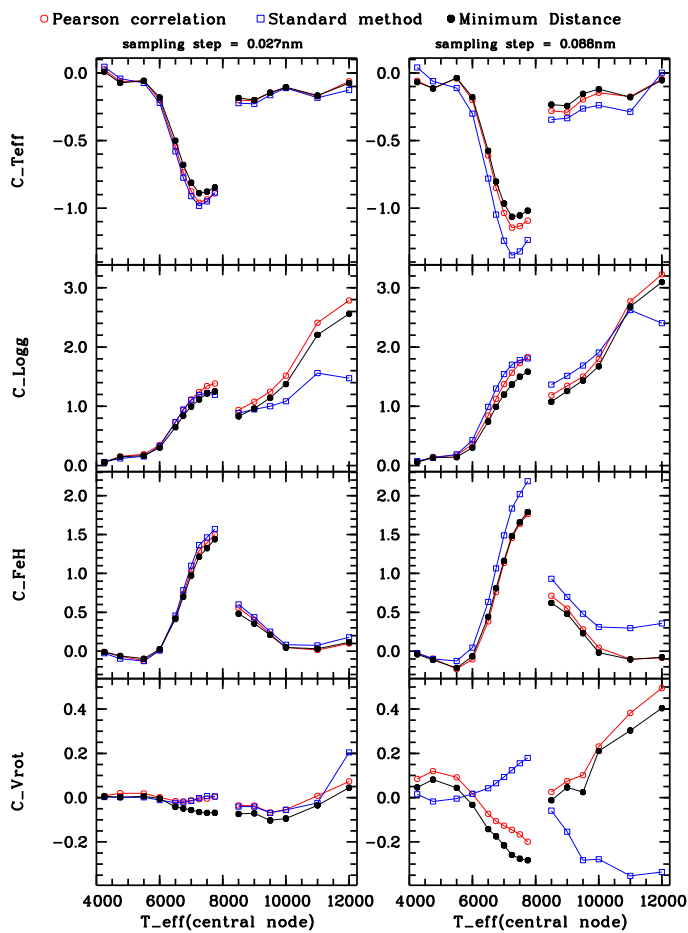}
\caption{The first-order coefficients in Eq.\,(\ref{eq:2nd}) for a set of 
central nodes with $\log g=4$, [Fe/H]$=0$, and $v\sin i=20$, but with 
different $T_\mathrm{eff}$-values as indicated on the x-axis. The y-axis 
labels identify the deviating AP in each frame. The ordinate values 
represent the TME-contribution (in $\kms$) of this AP if it deviates by 
one step (see Eq.\,(\ref{eq:apdev})) from the central node. The run of the 
coefficients is given for slightly oversampled (left frame) and 
slightly undersampled (right frame) spectra with an instrumental 
resolution of 11500, and for the three measurement methods as indicated on 
top. The gap at 8000\,K is due to the fact that different 
atmospheric models were used for temperatures above and below this value; incidentally it illustrates the sensitivity of Doppler-shift measurements 
to the templates used. 
}
\label{fig:c1}
\end{figure}

\section{Random errors }\label{sec:rerr}

\subsection{Internal error estimators }\label{sec:rest}
Besides the various algorithms for Doppler-shift measurement, also several 
ways of estimating the random error on this measurement have been proposed; 
the term {internal error} is borrowed from \citet{TODA79}. 
Most of these were designed in accordance with a specific C-function  
and with certain assumptions about the data, so some care is due 
if one wishes to apply them in a different context. 
We now discuss their applicability under the particular requirement that 
they be realistic (at least in order of 
magnitude) as well as suitable for a wide range of spectral structures. 

\citet{TODA79} estimated the error on the radial velocity using a measure 
of the asymmetry of the CCF peak, based on the argument that this peak is 
intrinsically symmetrical if the two spectra are noiseless and match each 
other exactly. Their estimate thus describes not only the random error but 
also (at least part of) the systematic error caused by spectrum mismatch. 
However, \citet{VEDA99} demonstrated that it cannot be valid for spectra
that exhibit relatively broad features such as the Balmer lines in the 
visible region of early-type spectra, the CaII triplet or the Paschen 
lines in the far red etc.  
  
\citet{MUHE91} proposed an error estimate based on the standard expression 
for the error on the least-squares estimate applied in centroiding the 
C-function peak. However, in doing so they presumed that the random error 
on neighbouring C-function samples is uncorrelated whereas actually it is 
highly correlated in general \citep{JEWA69}. In the case of the PCF and of 
the minimum-distance C-functions, the random error is in fact almost identical 
for adjacent C-function samples if these are calculated with a velocity 
step that is much smaller than the one corresponding to the step of the 
observed wavelength grid. Therefore this error estimate is not applicable 
here. 

\citet{BROW90} derived a lower bound for the random error, assuming that 
there is only photon noise. However, in most applications a realistic 
error estimate, rather than a mere lower bound is required. We verified in 
a number of cases that Brown's estimator indeed yields values that are 
too small to be relevant in practice.  

After extensive tests we found that the best error estimator to be used with 
both the CCF and the PCF, is the one proposed by \citet{ZUCK03}: 
\beq
\sigma_v^2 = \frac{1-C^2}{NCC^{\prime\prime}}  \ ,      \label{eq:sigcc}
\eeq
where $N$ is the number of wavelength bins while $C$ and $C^{\prime\prime}$ 
represent respectively the functional value and the second derivative of
the C-function ($C_\mathrm{cc}$ or $ C_\mathrm{pc}$), 
evaluated at its centroid position. 

For the minimum-distance functions we follow the approach of \citet{CASH79}. 
Each of these C-functions equals (up to some constant terms which have been 
omitted) the statistic $-2\ln L(a,v|f)$ and it can be shown that the 
difference $\Delta C$ defined as 
\beq
\Delta C(v) = C(v) -C(\hat{v})  \ ,                  \label{eq:delC}
\eeq
where $C$ stands for any of the functions $C_\mathrm{md}(v)$, 
$C_\mathrm{mdp}(v)$ or $C_\mathrm{mdc}(v)$ (see 
Appendix~\ref{sec:fldist}), is distributed approximately 
as $\chi^2$ with one\footnote{Even in the case where a is also adjustable: 
in fact we are not interested in the joint confidence region for both 
parameters but only in the overall uncertainty on $v$ \citep{PRFT86}.}
degree of freedom. Thus we can define a classical ``one-$\sigma$'' 
confidence region as the interval $[v_1,v_2]$ where 
$\Delta C(v_1) = \Delta C(v_2) = 1$. This interval needs not be 
symmetrical with respect to $\hat{v}$, so we adopt 
$\sigma_v = \max|v_i-\hat{v}|$ 
as the internal error estimate for a minimum-distance measurement. 

\subsection{Very faint objects}\label{sec:vfo} 

The internal error estimators all use (one way or another) the shape 
of the C-function in the immediate vicinity of its maximum. However, as the 
continuum $S/N$ decreases, at some level the small-scale behaviour of the 
C-function becomes dominated by the noise so that the position of the 
``highest peak'' may no longer be related to the actual Doppler shift, 
possibly resulting in a very large measurement error that cannot be 
predicted by the above estimators. This is illustrated in the bottom row 
of Fig.\,\ref{fig:amb} where the extremum values for the early-type star 
obviously do not indicate the correct Doppler shift. 
\begin{figure*}
\includegraphics[width=2\columnwidth]{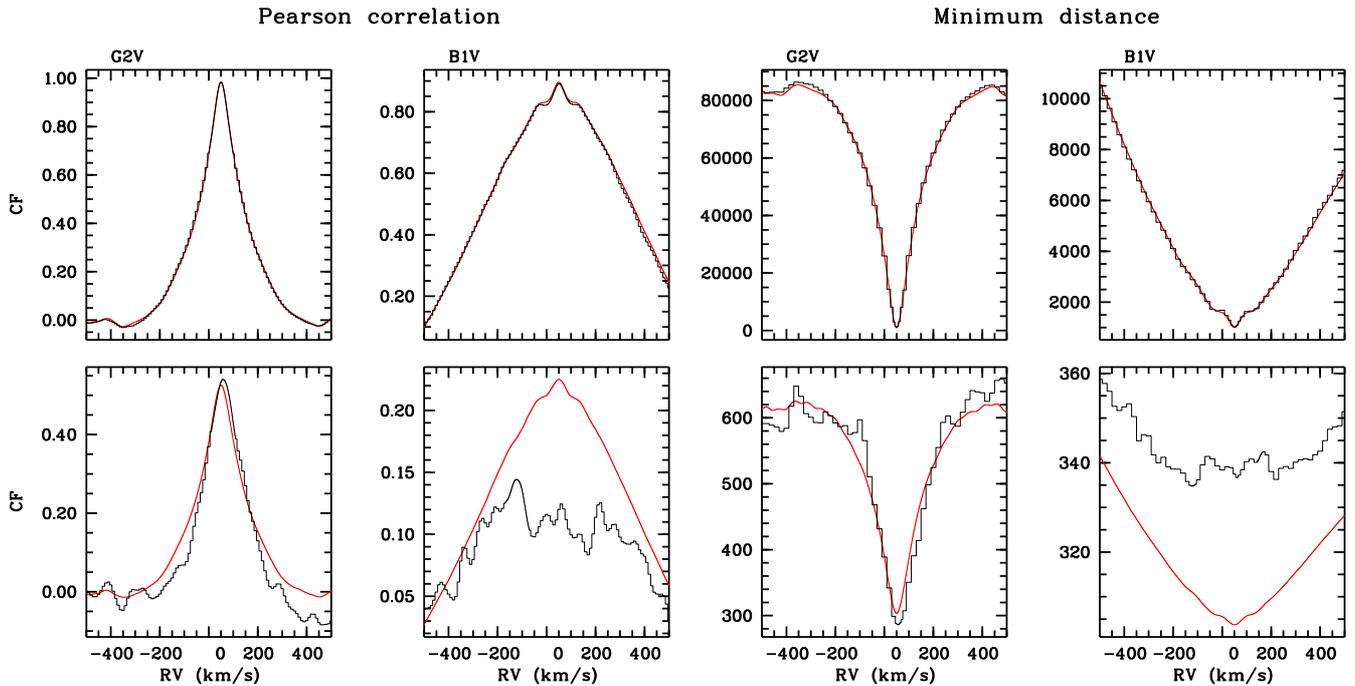}
\caption{C-functions on the wavelength range 847-874\,nm, obtained with 
two of our methods for a late-type and an early-type object as indicated 
(synthetic spectra with only photon noise, no rotational broadening 
and ``source'' velocity = $50\kms$).  
Top row: spectrum grid step = 0.027\,nm, continuum $S/N$ = 60; bottom row: 
spectrum grid step = 0.088\,nm, continuum $S/N$ = 7. The red lines indicate 
the average over a large number of C-functions obtained from spectra 
differing only by their noise realization.   
}
\label{fig:amb}
\end{figure*}

This phenomenon is well-known in signal analysis where it is sometimes 
referred to as ``ambiguity''. If the signal has a simple shape such as  
a radar pulse, the $S/N$ level that marks the onset of ambiguity can be 
predicted theoretically \citep{WODA50} but unfortunately a stellar spectrum 
containing lines of different widths and strengths, many of them blended, 
allows no such prediction; this is illustrated by a comparison of the 
columns 1 and 2 or 3 and 4 in Fig.\,\ref{fig:amb}: for the early-type 
spectrum we find significant ambiguity already at $S/N$ = 7, whereas 
for the late-type object this would occur at $S/N$=3 and 
below\footnote{It is appropriate here to point out that the quantity 
`continuum $S/N$' that we use as an indicator of apparent brightness,  
corresponds to the mean flux level (i.e. \# photons per wavelength bin) 
at the continuum. Therefore it should not be interpreted as a direct 
indication of the precision of information obtained from the spectrum.  
A realistic $S/N$ value in the latter sense (e.g. the mean ratio of the 
depth of line-cores to the amplitude of the noise) could be 2-10 
times smaller. }. 

Notice in particular the difference in the C-function profiles between the 
``cool'' object and the ``hot'' one: in the former the large-scale behaviour 
is dominated by the CaII lines while the relatively narrow central part is 
produced by the numerous metal lines in the spectrum; the early-type spectrum, 
on the other hand, contains only a small number of narrow lines which
produce a weak peak superposed on a very broad base dominated by the Paschen 
lines: this weak structure may yet allow a fairly accurate measurement if 
the $S/N$ is sufficiently high, but it is easily obliterated otherwise. 

Notice also that the C-function curvature at a typical ``pure noise''
extremum is not very different from the one at a proper peak; this partly 
explains why the random error is often badly underestimated in the case 
of ambiguity.

\subsection{The Monte-Carlo error bar}\label{sec:mce}

In spite of the best possible efforts, one may be faced with random error 
sources (such as ambiguity) whose effect cannot be described by the estimators 
above. A possible way of studying such errors is to use the following 
Monte-Carlo approach. For a given instrument and wavelength sampling one 
considers a 
wide variety of spectral-type, $v\sin i$, and $S/N$ combinations; for each of 
these one simulates a large number ($N_\mathrm{mc}$) of observations by 
adding noise to a spectrum with known source velocity; measuring the 
Doppler shift of these simulated spectra one obtains a sample of 
$N_\mathrm{mc}$ simulated errors: 
\beq
e_i = v_{i,\mathrm{meas.}}-v_{i,\mathrm{known}}  \ .       \label{eq:emeas}
\eeq 
The dispersion of these differences constitutes an alternative estimate of 
the random error, which we shall henceforth refer to as the 
{Monte-Carlo error bar} $\sigma_\mathrm{mc}$. A robust estimate of
this dispersion is obtained as the sample semi-interquantile range 
\citep{HOHO09}
\beq
\sigma_\mathrm{mc} = \frac{1}{2}
\left(\hat{Q}_{e}(0.8413)- \hat{Q}_{e}(0.1587)\right)    \label{eq:mcerr}
\eeq
that corresponds to the standard deviation if the simulated errors follow 
a normal distribution. 

The quantity $\sigma_\mathrm{mc}$ is not intended to replace an individual 
error estimate (except when the latter is obviously invalid) but it may 
provide useful additional information on the large-scale variation of 
random errors as a function of the atmospheric parameters of the observed 
source. Therefore, if one intends to observe a wide variety of objects 
with a given telescope-instrument combination, it is worthwhile again 
to create an auxiliary database from which an approximate value of 
$\sigma_\mathrm{mc}$ for any observed object can be obtained by interpolation. 
This can be done by calculating $\sigma_\mathrm{mc}$ for the nodes on 
a coarse grid such as the one described in Sect.\,\ref{sec:glomod}, but with 
an additional dimension corresponding to the $S/N$ level so that (unless 
additional APs need to be considered) it is five-dimensional.

\section{Testing the methods}\label{sec:tests}  
\subsection{The bias test}\label{sec:bias} 
Bias in a Doppler-shift measurement may originate from several sources. Most 
of these are best detected by performing a measurement on many different 
noiseless spectra for which (within the limits of numerical accuracy) the 
exact outcome is known, looking for 
deviations from the latter that are large enough to violate one's 
accuracy requirements. However,  one cannot 
exclude the possibility of some mechanism interfering with the propagation 
of random errors and causing the final error on the measurement to be 
biased. To test whether a given measurement technique is liable to such a 
bias, one can use again the set of  Monte-Carlo simulated errors of 
Eq.\,(\ref{eq:emeas}) and determine their median as a robust 
estimate of the bias. 

The standard deviation of the median can be estimated as  
 $\sigma_\mathrm{|M|} = \sigma_\mathrm{mc}1.2533/\sqrt{N_\mathrm{mc}}$ 
\citep[see e.g.][Example 10.7]{KEST69}. Then, from a two-tailed test one 
may conclude, at the significance level $\alpha$, that the 
measurement is biased if 
\beq
  \frac{|M|\sqrt{N_\mathrm{mc}}}{1.2533\sigma_\mathrm{mc}}                                           > F^{-1}(1-\frac{\alpha}{2})   \ ,              \label{eq:varM}
\eeq
where $F$ is the standard normal cumulative distribution function.

\subsection{The zscore test}\label{sec:zsc}

In principle precision can be assessed by means of the individual 
estimators discussed in Sect.\,\ref{sec:rest}, but naturally one needs 
to verify to what extent these are reliable for the objects one intends 
to observe; this can be done as follows.
We consider again a set of $N_\mathrm{mc}$ simulated errors 
(see Eq.\,(\ref{eq:emeas})) and denote the random-error estimate 
accompanying each measurement by $\sigma_i$; then we define the 
quantity 
\beq
z_i = \frac{e_i}{\sigma_i} \ ,                          \label{eq:zsc1}
\eeq 
which is termed the {zscore} of the measurement. If in fact the 
estimator correctly describes the true measurement error, the quantities 
$z_i$ follow a standard normal distribution. This could be ascertained by 
means of e.g. the one-sample Kolmogorov-Smirnoff test, but a quicker (and, 
as we found, sufficiently reliable) approach consists in determining their 
dispersion $\sigma_z$ and comparing this to its sampling distribution.  
Again we use the robust estimate
\beq
\sigma_z = \frac{1}{2}
\left(\hat{Q}_{z}(0.8413)- \hat{Q}_{z}(0.1587)\right)\ .   \label{eq:zsc2}
\eeq 
The sampling distribution of $\sigma_z$ for a standard normal variable is 
normal with mean=1 and standard deviation 
$\mathrm{stdev}[\sigma_z] \approx {0.962}/\sqrt{N_\mathrm{mc}}$ 
\citep[see e.g.][Example 10.8 with $p_1=0.8413$ etc.]{KEST69}. Then, from 
a two-tailed test one may conclude, at the significance level $\alpha$, 
that the estimator is not valid if 
\beq
\frac{|\sigma_z-1|\sqrt{N_\mathrm{mc}}}{0.962} > F^{-1}(1-\frac{\alpha}{2}) 
                                           \ .          \label{eq:zsc3}
\eeq

\subsection{Comparing performances }\label{sec:perf} 

Consider any two measurement methods that both satisfy the above quality 
requirements; then naturally the question arises whether 
one of them nevertheless induces somewhat larger errors than the other and, 
more to the point, under what circumstances this is the case. To investigate 
this we consider, again, sets of Monte-Carlo simulated errors as defined 
in Eq.\,(\ref{eq:emeas}), pairing the samples so that we always compare two 
measurements obtained with the same noise realization. 
Although we shall frequently use the terms ``better'' or 
``best'' in the following discussions, this will never be meant  
in an absolute sense, firstly because we only consider the magnitude of 
the errors (disregarding e.g. the computational load of a given method)  
and secondly because the test results are likely to depend on resolution 
and sampling, on the noise level, the spectral type, and the rotational 
broadening of the objects. 
A different choice for these parameters could easily lead to a 
different conclusion. 

Supposing one wishes to test separately for differences in bias and in 
dispersion, one could proceed as follows. In a set of Monte-Carlo 
simulated errors the bias appears as the mean of these errors; a classical 
test to detect a difference in the means of two normally distributed 
populations is based on Student's $t$-distribution and it has a version 
that is applicable to paired samples. However, for the dispersion the 
situation is different: we found no paired-sample counterpart for the 
classical F-test on the homogeneity of variances; therefore we propose to 
use again the $t$-test, but now on the absolute value of the errors after 
subtraction of the bias, as detailed below.    

These are not the only possibilities of course. If one is interested 
e.g. in finding out whether there is a significant difference in the 
{total} error (bias + random part) one could repeat the second test 
without subtracting the bias. And if one is worried about the fact that a 
difference of absolute values of errors may not be normally distributed, 
one could replace the $t$-test by a non-parametric one such as the 
{Sign test} \citep[see e.g.][]{SIEG56}.  

\subsubsection{Bias}

Let us name the methods, arbitrarily, ``1'' and ``2''; we note $e_{1,i}$ 
and $e_{2,i}$ for the respective error incurred with each of the 
$N_\mathrm{mc}$ simulated spectra and we define the differences 
\beq
d_i=e_{1,i} - e_{2,i}   \ .                           \label{eq:di}
\eeq
Next we compute the average $\overline{d}$ and the standard deviation 
$\sigma_{\rm d}$. If both methods perform equally well as far as the 
bias is concerned, the {population mean} of $d$ equals zero; 
furthermore, provided that both $e_{1,i}$ and $e_{2,i}$ follow a 
Gaussian distribution with the same dispersion, the statistic 
$t_{\rm d} = \overline{d}/(\sigma_{\rm d}/\sqrt{N_{\rm mc}})$ is 
distributed according to a Student's $t$-distribution with $N_{\rm mc}-1$ 
degrees of freedom \citep[see e.g.][]{ALRO72}. If $N_{\rm mc}$ is 
sufficiently large (e.g. $\ge 500$) the $t$-distribution  may be replaced 
by the standard normal.  Then, from a two-tailed test one concludes, 
at the significance level $\alpha$, that this null-hypothesis must be 
rejected if 
\beq
\frac{\overline{d}\sqrt{N_{\rm mc}}}{\sigma_{\rm d}}
 > F^{-1}(1-\frac{\alpha}{2})               \ .              \label{eq:stut}
\eeq
Subsequently a comparison between $|\overline{e_{1}}|$ and 
$|\overline{e_{2}}|$ will indicate which of the two methods causes the 
smallest bias. Note that, 
if $\overline{e_{1}}$ and $\overline{e_{2}}$ have different signs, the 
statistic may indicate a significant difference without either of the 
methods being truly ``better'' than the other. Also, even if they have 
the same sign and a statistically significant difference, it is possible 
that this difference is not {physically} relevant because the largest 
bias is still sufficiently small to be harmless for the required accuracy.

\subsubsection{Dispersion}
For the present purpose we use the {mean absolute deviation} to 
characterize the dispersion. Let \\
$e'_{1,i} = e_{1,i} - \overline{e_1}\ ,\ \  
e'_{2,i} = e_{2,i} - \overline{e_2}\ ,\ \  
d'_i = |e'_{1,i}| - |e'_{2,i}|$ 
and compute the average $\overline{d'}$ and the standard deviation 
$\sigma_{\rm d'}$. If both methods produce measurements with the same 
dispersion, we expect that the {population mean} of $d'$ equals zero. 
Furthermore we assume that the differences $d'_i$ are normally distributed 
so that the statistic 
$t_{\rm d'} = \overline{d'}/(\sigma_{\rm d'}/\sqrt{N_{\rm mc}})$
follows a $t$-distribution with $N_{\rm mc}-1$ degrees of 
freedom. Then, as for the bias-difference, we can conclude that the 
difference $\overline{|e'_1|}-\overline{|e'_2|}$ is significant at 
the level $\alpha$ if 
\beq
\frac{\overline{d'}\sqrt{N_{\rm mc}}}{\sigma_{\rm d'}}
 > F^{-1}(1-\frac{\alpha}{2})  \ ,                          \label{eq:stut'}
\eeq
and in that case, see which method causes the smallest mean absolute 
deviation.  

\subsubsection{Application}

\begin{figure*}
\centering
\resizebox{\hsize}{!}{\includegraphics{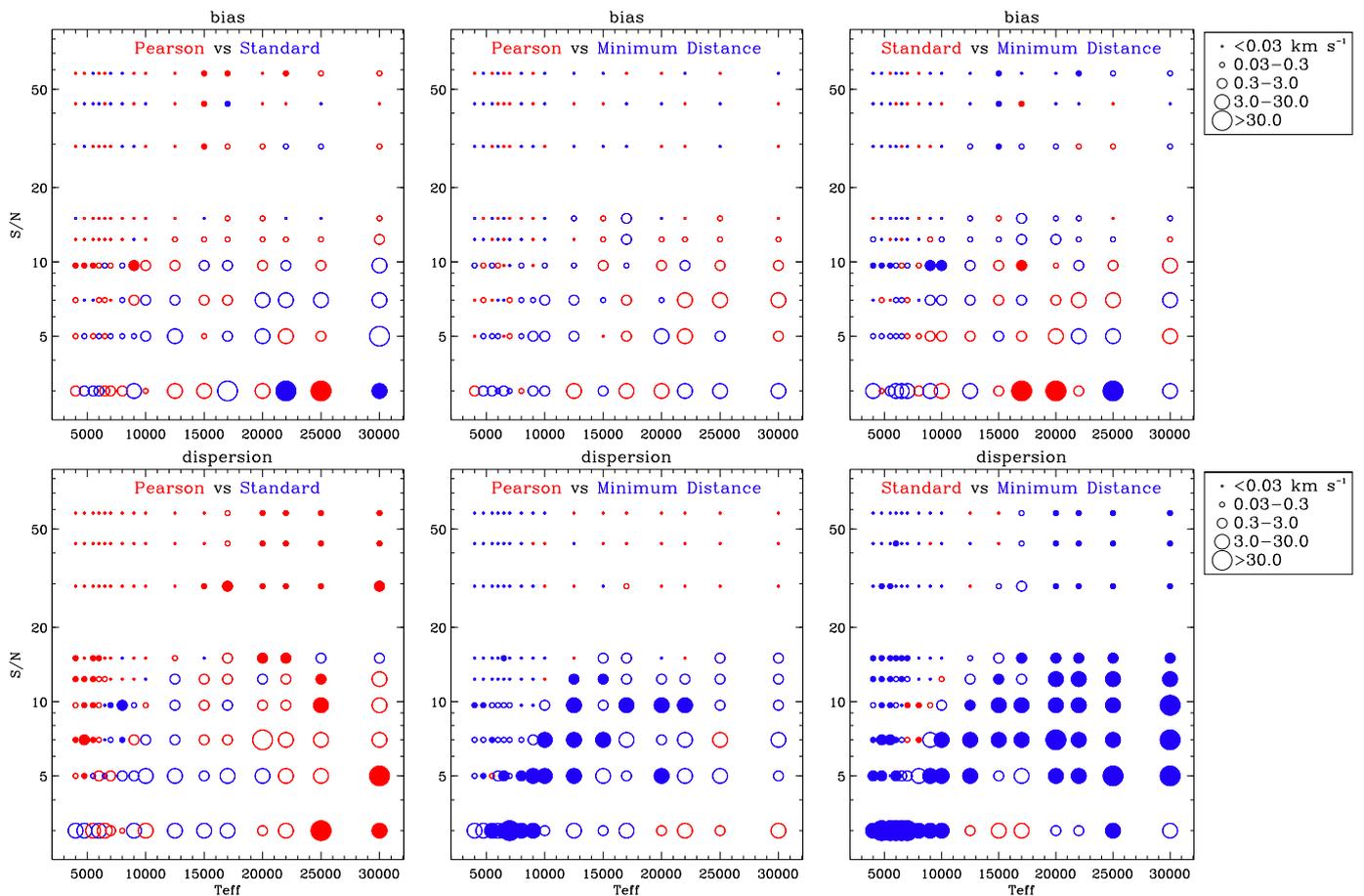}}
\caption{Pairwise comparison of the bias (top panels) and the
dispersion (bottom panels) between three different methods 
as indicated in each panel. 
The results are plotted as a function of effective temperature (x-axis)
and signal-to-noise ratio (y-axis).
The colours are identified with the method names in each panel; 
each point is given the colour of the method with the smallest bias 
(top row) or the smallest dispersion (bottom row) respectively. Filled 
symbols indicate that the difference is statistically significant at 
the 0.2\% level in a two-sided test. }
\label{fig:dispersion}
\end{figure*}

To illustrate the use of these tests, we apply them to a set of spectra 
corresponding to different values of effective temperature and 
continuum $S/N$ ratio;  
otherwise all spectra have $\log g = 4.0$, $v \sin i = 0.0\kms$, 
and solar abundances. 
The spectra simulate observations from the Gaia-RVS instrument: they cover 
the wavelength range 847--874\,nm with an instrumental 
resolution of 11,500; the wavelength sampling-step is 0.027\,nm for the 
brighter spectra (continuum $S/N>25$) and 0.088\,nm otherwise. 
For each combination of $T_\mathrm{eff}$ and $S/N$, a sample of 
$N_{\rm mc}=1000$ spectra was generated with different noise realizations.  

In Fig.\,\ref{fig:dispersion} we compare the performance of the methods two 
by two. In order to be able also to judge the 
{physical relevance} of the differences, we scaled the size of the 
symbols with the difference in bias magnitude   
\beq
s_\mathrm{d} = \left||\overline{e_1}| - |\overline{e_2}|\right|
                                                          \label{eq:sd}
\eeq
or with the difference in standard deviation 
\beq
s_\mathrm{d'} = \left|\mathrm{stdev}(e_1) - \mathrm{stdev}(e_2)\right|\ ,
                                                          \label{eq:sd'}
\eeq
both expressed in $\kms$. 
Note that larger differences are not necessarily also {statistically} 
more significant. 

For these specific data, there are clear areas in the ($T_\mathrm{eff}$, 
$S/N$) diagram where one technique provides a significantly 
sharper dispersion than the other.  Differences between the techniques also 
exist for the bias, but these seem to be less consistently grouped into 
areas of the diagram and they are, on the whole, statistically less 
significant. We caution again that the conclusions derived here apply to the 
specific instrument for which these simulated data were calculated. Other 
instrument configurations (wavelength range, resolution, ...) could easily 
lead to different conclusions. 


\subsection{Global prognostics for a survey}\label{sec:adb} 

The various tests described above do not involve observed data so they can 
be performed independently of the actual measurements, though for a given 
telescope-instrument combination. Thus, if one expects to observe a wide 
variety of spectra it is best to perform them as early as possible 
and to store their results in an auxiliary 
database (ADB) such as those already mentioned in Sects.\,\ref{sec:glomod} and 
\ref{sec:mce}. 

In a multi-method approach, all measurement algorithms are naturally 
implemented on a common platform where most of the peripheral operations 
(e.g. data preparation and template construction) are performed by separate modules communicating with the several measurement modules through 
appropriate interfaces. Such an architecture favours the implementation of 
a dedicated "tests" module that 
\begin{itemize}
\item{}defines a common (coarse) AP test grid\footnote{e.g. as described 
in Sect.\,\ref{sec:mce}}; 
\item{}generates $N_\mathrm{mc}$ simulated spectra for each node; 
\item{}calculates the Monte-Carlo error bar 
$\sigma_\mathrm{mc}$ and the left hand side of Eqs.\,(\ref{eq:varM},\ref{eq:zsc3}) 
with each of the measurement methods;  
\item{}for each couple of methods, calculates the left hand side of 
Eqs.\,(\ref{eq:stut},\ref{eq:stut'}) and the size differences 
(Eqs.\,(\ref{eq:sd},\ref{eq:sd'})); 
\item{}stores all results in the ADB. 
\end{itemize}
It is convenient to calculate the coefficients of the local model 
for the TME (Eq.\,(\ref{eq:2nd})) on the same grid (but without the 
$(S/N)$ dimension) so they can be stored in the same ADB.   

The "tests" module should also offer appropriate tools for querying the 
ADB and for plotting, so that graphs such as Fig.\,\ref{fig:c1} or 
Fig.\,\ref{fig:dispersion} can be produced easily. Thus from 
this database one can obtain an overall picture of the expected 
quality of the Doppler-shift measurements. This in turn may provide 
useful feedback to further improve the implementation of the methods 
and/or to modify the observational set-up 
or even to avoid observations that are likely to be profitless.  

\section{Combining the measurements }\label{sec:comb}

The user of any measurement software naturally wants a single answer  
within some confidence region. Assume now that we have an auxiliary 
database as described above and that a (common) significance level 
$\alpha$ for the various hypothesis tests has been chosen; then, for 
each observed spectrum we can proceed as follows:  
\begin{itemize}
\item{}determine which node on the grid best matches the observation and 
retrieve the corresponding data to be used in the following steps;
\item{}select the methods that pass the bias test; the measurement 
produced by a method that fails this test should be stored but 
not used in a combination; 
\item{}if any of the remaining methods fails the zscore test, its internal
error estimate must be replaced by its Monte-Carlo error bar; 
\item{}considering the TME as a random effect (see Sect.\,\ref{sec:tmgen}), 
estimate its magnitude\footnote{a safe option is to take the maximal 
magnitude of the TME over the unit cell (centred on the 
object's APs) of the fine grid defined in Sect.\,\ref{sec:grid}}
using the local-model coefficients for 
each remaining method and add it quadratically to the random error; 
\item{}obtain the combined measurement as the median or the error-weighted 
average of the remaining individual measurements; 
\item{}estimate the combined error using standard error propagation;
\end{itemize} 
The above procedure is not very refined, but at least its 
automation is feasible and it should provide more reliable answers than 
a simple average of all measurements. Obviously 
its success will hinge on the richness and the quality of the auxiliary 
database. 

Nevertheless one should remain aware of the fact that there may be 
situations (e.g. when the various measurements {all} disagree within the 
limits of their individual uncertainty, as is not inconceivable with faint 
objects) where one could hardly have confidence in the result of any 
automated combination, however sophisticated be its scheme. 
Therefore it is best to keep on record not only the 
combined value and its error estimate, but also the result produced by each 
method separately. In this way the user can judge whether it is necessary 
to have a closer look at the individual results and perhaps combine these 
in another way or to go back even further and handle the measurement 
manually for the case at hand.

\section{Conclusions}\label{sec:con}

Among the numerous methods for Doppler-shift measurement available in the literature, we select three that do not impose strong requirements on the 
observed data, regarding resolution and $S/N$. For one of these we provide 
two variants that one may use lacking a realistic estimate of the error 
on each flux sample. The theoretical 
foundation of the methods is discussed to provide a better understanding of   
any differences in the results they yield. 

With each method we select an appropriate estimator for the internal 
error on the measurements and we propose a Monte-Carlo based global estimator, 
to be used e.g. with very faint objects where the internal estimator 
becomes unreliable. We also discuss in  some detail how the so-called 
template mismatch error can be dealt with in a systematic way.
Several tests allowing to judge the performance 
of the methods are described. 

In our multi-method approach we propose a combination of the individual 
Doppler-shift measurements, using an auxiliary database that roughly 
predicts the measurement-quality one can expect for any type of object 
to be observed. If such a 
database can be established already at an early stage in a survey project, 
it may moreover provide useful information to optimize the instrumental 
set-up and the observational strategy or even the instrumental design.

\begin{acknowledgements}
We acknowledge funding by the Belgian Federal Science Policy Office
through ESA PRODEX programme ``Binaries, extreme stars and solar 
system objects'', under contract nrs. C90289 and C90290.
Work done in GEPI has been partly funded by CNES through programme 
``CNES-Gaia" under contract nr. 92532/o/. 
The authors would like to thank the Gaia Data Processing and Analysis Consortium, 
and in particular the colleagues in Coordination Unit 6
(``Spectroscopic Processing''), for numerous fruitful 
discussions.  

\end{acknowledgements}

\begin{appendix}

\section{Template sampling in $\ln{\lambda}$ space}\label{sec:tmpls}

If the wavelength grid has a constant step $\Delta$ in 
$\ln{\lambda}$, the central wavelength of the bins can be written as  
\beq
\lambda_n = \lambda_0\,e^{n\Delta}   \ .                    \label{eq:clnl}
\eeq 
Since the sampling Eq.\,(\ref{eq:giv}) is contiguous we have 
$\lambda_n + \Delta_n = \lambda_{n+1} - \Delta_{n+1}$ so that, noting 
$\mu_n=\lambda_n - \Delta_n$ we can write $\lambda_n + \Delta_n = \mu_{n+1}$ 
and thus $\lambda_n=\frac{\mu_n+\mu_{n+1}}{2}$; then it is easily seen that  
\beq
\mu_n = \mu_0e^{n\Delta}\ \ ,\ \ \ \mu_0  = 
\frac{\lambda_0\,e^{-\frac{\Delta}{2}}}{\cosh(\frac{\Delta}{2})} 
                                                      \ .    \label{eq:lmd}
\eeq 
Consider now in particular the discrete set of velocities satisfying 
\beq 
1+\frac{v_m}{c}=e^{m\Delta}\ \ ,\ \ \ m\in\NN   \ ;          \label{eq:vis}
\eeq
then, provided the approximation Eq.\,(\ref{eq:giv1}) is valid: 
\beqa
t_n(v_m)&=&\int_{\mu_ne^{-m\Delta}}^{\mu_{n+1}e^{-m\Delta}}
t(\lambda;0)\dff\lambda  =  \int_{\mu_{n-m}}^{\mu_{n-m+1}}
t(\lambda;0)\dff\lambda                  \nonumber    \\                   
&=& t_{n-m}(0)                           \ .         \label{eq:dsh}
\eeqa 
Thus we obtain the familiar picture of a rigid shift of the spectrum as 
a result of the motion of the source. This implies in particular that 
for different velocities satisfying Eq.\,(\ref{eq:vis}), even the 
simplified sampling in Eq.\,(\ref{eq:giv1}) 
does not have to be recalculated for each velocity.  

If other velocities need to be considered, it may be useful to rewrite 
Eq.\,(\ref{eq:giv}) in a form that closely resembles Eq.\,(\ref{eq:dsh}) 
by defining a new variable $s$: 
\beq 
1+\frac{v}{c}=e^{s\Delta}                               \label{eq:vrs}
\eeq
so that $t_n(v)$ can be written as a function of $n-s$: 
\beq
t_n(v) = h(n-s) = \int_{\mu_0e^{(n-s)\Delta}}^{\mu_0e^{(n+1-s)\Delta}}
t(\lambda;0)\dff\lambda        \ .                \label{eq:grv}
\eeq 
Thus Eq.\,(\ref{eq:vrs}) defines the Doppler shift as a real quantity, rather 
than an integer as in Eq.\,(\ref{eq:vis})

\section{Other flux distributions}\label{sec:fldist}

If a source is sufficiently bright one can assume that the PDF is Gaussian 
and estimate its dispersion as $\sigma_n^2=at_n(v)+\sigma_\mathrm{ro}^2$ 
where $\sigma_\mathrm{ro}$ represents e.g. ``read-out noise'' or any constant 
additive noise source. Note that, unlike $\sigma_n$ in Eq.\,(\ref{eq:gpdf}), 
this error model depends on the template velocity. Thus, instead of 
Eq.\,(\ref{eq:gpdf}) we now have to minimize  
\beq
C_\mathrm{mdp}(a,v) = \sum_{n=1}^N \left\{\frac{(f_n-at_n(v))^2}
{at_n(v)+\sigma_\mathrm{ro}^2} + \ln(at_n(v)+\sigma_\mathrm{ro}^2)\right\} 
                                    \ .                     \label{eq:gphro}
\eeq
An exact 1-dimensional 
version of this function has not been found. However, if the scale factor 
$a$ is known independently we can put $a=1$ and then, neglecting the 
logarithmic term and assuming there is only photon noise, Eq.\,(\ref{eq:gphro}) becomes the so-called {S-statistic} \citep[see e.g.][and references 
therein]{CASH79,LAMB76}. 
It should be noted that the error model assumed above cannot be used down to 
arbitrarily low flux levels. In such cases one would have to build a somewhat 
more sophisticated PDF that properly accounts for the 
combination of photon noise and an additional Gaussian noise source. 

If all noise sources except photon counting are negligible, 
the flux-sample values are Poisson-distributed, i.e.   
\beq
\mathrm{pd}(f_n|a,v) = \frac{(at_n(v))^{f_n}e^{-at_n(v)}}{f_n!}
                                             \ .              \label{eq:popd}
\eeq
In that case, maximizing the likelihood is equivalent to minimizing 
\beq
C_\mathrm{mdc}(a,v) = 2\sum_{n=1}^N \left\{at_n(v) - f_n\ln(at_n(v))\right\} 
                                          \ ,               \label{eq:Po}
\eeq
where the factor 2 is introduced to allow for the application of 
Eq.\,(\ref{eq:delC}). At the extremum we now have   
\beq
\hat{a} = \frac{\sum f_n}{\sum t_n(\hat{v})}   \ ,          \label{eq:hapo}
\eeq
which can be substituted in Eq.\,(\ref{eq:Po}) to obtain a 1-dimensional 
C-function. On the other hand, if again the parameter $a$ can be 
omitted, the right hand side of Eq.\,(\ref{eq:Po}) equals the so-called 
{C-statistic} \citep{CASH79}. 
The C-function from Eq.\,(\ref{eq:Po}) is especially 
suitable when the observations deliver a pure photon count of a very weak 
source, without any other noise sources to be considered.  

From a performance-comparison as in Fig.\,\ref{fig:dispersion} we found 
that $C_\mathrm{md}$ is preferable but that both $C_\mathrm{mdp}$ and 
$C_\mathrm{mdc}$ are viable alternatives if, for some 
reason, $C_\mathrm{md}$ cannot be used.    

To the best of our knowledge, neither of the functions $C_\mathrm{mdp}$ 
and $C_\mathrm{mdc}$ has been used before in radial 
velocity measurements. We note, however, that the problem of Doppler-shift 
measurement is mathematically very similar to the detection of periodicity  
in time-series analysis, for which the {\sc mlp} method \citep{ZHCZ02} does 
use the C-statistic. 

\section{Upper bound for the MME}
\subsection{Symmetric functions}\label{sec:ubsym}

We note $x$ for the independent variable, $y$ for the corresponding 
function value, $s$ for the sampling step size, $(x_0,y_0)$ for the 
data point with the highest/lowest sample value within the range of $x$ 
and $(x_0\pm s,y_\pm)$ for the nearest adjacent samples so that
\begin{equation}
(y_0 - y_+)(y_0 - y_-) \geq 0       \ .                      \label{eq:ineq}
\end{equation}
Furthermore we note $x_\mathrm{c}$ for the true position of the extremum 
and $x_\mathrm{p}$ for the peak-position estimated using a parabola as 
interpolating function. Then   
\begin{equation}
x_\mathrm{p} = x_0 - \left(\frac{y_+ - y_-}{y_+ + y_- - 2y_0}\right)\frac{s}{2} 
                               \ .                           \label{eq:ext}
\end{equation} 
Even if the function is intrinsically symmetric this estimate has a modest 
MME related to the discrete nature of the sampling, i.e. there is a 
difference    
\begin{equation}
\delta = x_\mathrm{p}-x_\mathrm{c}                        \label{eq:mme}
\end{equation}
that depends on the actual shape of the function. It is easily seen that  
$\delta = 0$ in the special cases $y_+ = y_-$ and $y_0 = y_\pm$; otherwise, 
assuming that the function has no structure on a scale less than $s$ and 
that there is no noise, one finds that 
$|\delta | < \frac{s}{2}$. Of course the error will be much less than this if the function closely resembles a parabola. 

Anyway, it follows that $|\delta|$ could be made arbitrarily small by 
reducing the step size, as is possible with the velocity-space methods in 
Sect.\,\ref{sec:CFS}. However, in the 
standard method the step size $s$ cannot be reduced at will so there we 
reduce the basic MME by combining the above three-point fit with a 
four-point fit following \citet{DAVE95}. 

\subsection{Weakly asymmetric functions}\label{sec:ubasy}

We assume that $s$ is sufficiently small to ensure that, 
within the interval $[x_0-2s,x_0+2s]$, the function to be centroided 
is well approximated by a third-degree polynomial:  
\begin{equation}
y(x) = \alpha + \beta(x-x_\mathrm{c})^2 + \gamma(x-x_\mathrm{c})^3 
                                                            \label{eq:cfu0}
\end{equation}
where the parameter $\beta$ can be said to control the contrast of the 
extremum value with respect to the adjacent points and $\gamma$ to control 
the asymmetry\footnote{Of course this interpretation is meaningful only if 
$|\gamma s| << |\beta|$; as an example, for the red lines in the top 
row of Fig.\,\ref{fig:amb}, one has 
$5.10^{-4}< |\frac{\gamma s}{\beta}|< 10^{-2}$.}; 
if Eq.\,(\ref{eq:cfu0}) represents a C-function, $1/\beta$ reflects the width 
of the narrowest lines present in the spectra \citep[see also][]{VEDA99}.

Then, with the notations introduced above, choosing $x_0$ according to 
(\ref{eq:ineq}), after some algebra one finds that   
\beq
|\delta| < \frac{\left|\frac{\gamma s}{\beta}\right| s}
{1 - 3\left|\frac{\gamma s}{\beta}\right|}  
\ \ \ \mathrm{if}\ \ s < |\frac{\beta}{3\gamma}| \ .         \label{eq:maxe}
\eeq
\end{appendix}

\bibliographystyle{aa}
\bibliography{RV}

\end{document}